\documentclass[12pt,a4paper]{article}
\usepackage{graphicx}

\begin{document}
\begin{center}
\textbf{Category:} Original article

{\Large \bf Cooperation between genetic mutations and phenotypic plasticity can bypass the Weismann barrier: The cooperative model of evolution}  

{Ken Nishikawa$^{a,*}$ and Akira R. Kinjo$^a$\\
$^a$Institute for Protein Research, Osaka University, \\
3-2 Yamadaoka, Suita, Osaka 565-0871, Japan}

\textbf{Running title:} The cooperative model of evolution

\textbf{Corresponding author:} ken-nishikawa@protein.osaka-u.ac.jp
\end{center}

\begin{abstract}
The Weismann barrier, or the impossibility of inheritance of acquired traits, comprises a foundation of modern biology, and it has been a major obstacle in establishing the connection between evolution and ontogenesis. We propose the cooperative model based on the assumption that evolution is achieved by a cooperation between genetic mutations and acquired changes (phenotypic plasticity). It is also assumed in this model that natural selection operates on phenotypes, rather than genotypes, of individuals, and that the relationship between phenotypes and genotypes is one-to-many. In the simulations based on these assumptions, individuals exhibited phenotypic changes in response to an environmental change, corresponding multiple genetic mutations were increasingly accumulated in individuals in the population, and phenotypic plasticity was gradually replaced with genetic mutations. This result suggests that Lamarck's law of use and disuse can effectively hold without conflicting the Weismann barrier, and thus evolution can be logically connected with ontogenesis.

\end{abstract}

\begin{flushleft}
\end{flushleft}

\section{Introduction}
The modern ``evolutionary synthesis'' was established by integrating Darwin's theory of natural selection and genetics. However, at that time, the process of ontogenesis was considered as a black box, and hence was effectively ignored. In the 1940's when the theory of evolutionary synthesis was being established, developmental biology of the time was in the middle of classical biology that was full of mysterious phenomena. Therefore, it was understandable that such an under-developed field was ignored. Since then, however, developmental biology has been greatly modernized, owing to the huge success of molecular biology. In the 1980's, the discovery of Hox genes revealed the relationship between embryogenesis and genetic control \cite{Gehring1998}. Furthermore, in this century, epigenetics has become an active field of research in developmental biology.

Epigenetics refers to mechanisms of genetic control that are achieved by modification of chromatins (such as DNA methylation and post-translational modification of histones) without any change in the genetic information in the genome (i.e., the DNA sequence itself). Epigenetic chromatin modifications accompany cell differentiation, and the modifications are copied to daughter cells. Thus, the mystery of differentiation (how somatic cells can be different in spite of the same genome) has been resolved at the molecular level. Furthermore, epigenetics is affected by the environmental and nutritional conditions of individuals, leading to changes of phenotypes (e.g., obesity). External stresses in early stages of ontogenesis may result in malformations. Furthermore, many diseases, such as lifestyle related diseases and cancers, are caused not only by genetic factors, but also by its combination with environmental and acquired factors \cite{Francis2011}.

Natural selection operates on phenotypes of individuals rather than their genotypes. It does not act on genes themselves, but on various phenotypes ranging from molecular structures to individual body. Therefore, it is clear that, in evolution, we cannot ignore  the process of ontogenesis that includes epigenetic variations, which in turn, may affect phenotypes without modifying genotypes. In fact, the ``phenotype-driven mechanisms of evolution'' have been an active area of research in recent years, and these theories propose that a change in environment first affects the processes of embryogenesis and ontogenesis leading to changes of phenotypes, that in turn leads to evolution \cite{West-Eberhard2003,KirschnerANDGerhart2005,GilbertANDEpel2008}.

However, all attempts to relate evolution and ontogenesis have been rejected by the prohibition that ``acquired traits are not inherited.'' The prohibition is known as the Weismann barrier. More precisely, it states that germ cells are separated from somatic cells and there is no way to transmit information from the latter to the former. Naturally, this prohibition is included in the theory of evolutionary synthesis as an important principle. Since acquired traits are never inherited, it is unnecessary to consider any acquired elements including the process of ontogenesis, and hence only the mutations of genotypes, rather than of phenotypes, are sufficient for explaining evolution. This is how it is rationalized to ignore ontogenesis in evolutionary mechanisms based on the prohibition of inheritance of acquired traits. However, ignoring ontogenesis implies that ignoring natural selection on phenotypes. Thus, the standard approach of evolutionary synthesis is clearly unnatural and inevitably leads to difficulties in explaining, for example, the qualitative changes of traits accompanying adaptive evolution \cite{GilbertANDEpel2008}.

In the following, we present a model that can bypass the prohibition of inherited acquired traits without violating it. If this model is valid, the difficulty in connecting evolution with ontogenesis will be resolved, and the study of evolution will enter a new stage.

\section{Phenotypic mutation and genetic mutation}
The idea that phenotypic mutations drive evolution has been around for some time. The problem is how such mutations are inherited to next generations without breaking the Weismann barrier. In this respect, one plausible concept is ``genetic assimilation'' proposed by Schmalhausen \cite{Schmalhausen1949} and Waddington \cite{Waddington1953}, which is summarized as follows \cite{GilbertANDEpel2008}: ``If [...] plastic response is adaptive, and if it continues to be induced by the environment, it will spread under continued selection. Moreover, if this environmentally induced phenotype confers greater fitness, and if having the phenotype produced from the genome provides more fitness than acquiring it from the environment, then it should become stabilized by genetic means, if possible.'' To summarize this statement yet further, ``genes are followers, not leaders, in evolution'' \cite{West-Eberhard2003}.

The mechanism of genetic assimilation was proposed before the discovery of the double helix of DNA, and from the present standard, it appears too simplistic and somewhat out-dated. For example, it is assumed that one phenotypic trait is determined by one gene, or a single mutation changes a trait, etc. Complete sequencing of genomes has changed our view of genes completely. For example, the set of genes encoded in genomes are hardly different among morphologically very different mammals such as human and whale. This aspect is, of course, understandable if we consider that the variety of genes mainly corresponds to the cellular structures and functions, and that, furthermore, the repertoire of cells are hardly different between human and whale. Thus, it is now believed that macroscopic phenotypes are determined not by genes themselves, but by other subtle and diverse mechanisms that regulate gene expression \cite{Carroll2005}.

The subtle relationship between macroscopic phenotypes and (microscopic) genotypes discussed above can be also understood through more familiar human diseases. Diseases are a kind of phenotypic changes, and many of them, including diabetes, cerebral vascular disorder, cardiovascular diseases, and cancers, are considered to be multifactorial diseases each of which involves many genetic factors. As these examples show, the relationship between macroscopic phenotypes and genotypes are one-to-many whether the phenotypes are normal or abnormal. In addition, in many cases, one genetic factor causes multiple phenotypic traits.
Therefore, the relationship between phenotypes and genotypes is not as simple as it used to be considered, but is very subtle and complex. Regarding lifestyle related diseases, their outbreak depends not only on genetic factors, but also on the so-called environmental factors. Here, the environmental factors refer to all the external and acquired causes other than genetic factors, and they include, for example, diet, smoking, and other various stresses. When environmental factors act on a body, the body responds to it by the mechanism of phenotypic plasticity. For instance, improper diet and the lack of physical exercise may lead to obesity. It is also known that when one experiences insufficient nutrition before birth, he will become obese when becoming an adult \cite{Francis2011}.
Obesity itself is not a disease, but when combined with certain genetic factors, it will result in, say, diabetes. Since diabetes is a multifactorial disease, its outbreak depends on the number and kinds of genetic mutations. In the following, any variation in the genome sequence is called a ``genetic mutation'' by convention.

When we take into account the above mechanism of outbreak of diseases as well as the evolutionary mechanism by genetic assimilation, a new mechanism of evolution emerges, which is presented in the following.

\section{The cooperative model}
Suppose a species in a certain natural (ecological) environment. As far as the environment does not change, the species will only maintain the status quo, and hence will not evolve. In order for a species to evolve, there must be a change in environment that persists for a prolonged period. The species will try to adapt to the changed environment, but the extent of the adaptation will not be uniform in the population and varies between individuals. Individuals that are relatively well adapted to the environmental change are more likely to be naturally selected and to produce more offspring than those that are not so well adapted.

In order for a species to survive in a new environment, it has to express a phenotypic mutation that is adapted to the environment. There are three possible mechanisms for realizing adaptive phenotypic mutations:
\begin{description}
\item[Case A.] By genetic mutations.
\item[Case B.] By phenotypic plasticity.
\item[Case C.] By a cooperation between genetic mutations and phenotypic plasticity.
\end{description}
Case A corresponds to the conventional evolutionary synthesis, and it has a difficulty that the probability of adaptive genetic mutations to occur at the right time is very low. In case B, individuals with adaptive phenotypes will be selected, but without adaptive genetic mutations, such adaptive phenotypes will not be inherited to their offspring, and hence no evolution. Case C, as we have discussed above regarding lifestyle related diseases, comprises the essence of our model. Here, phenotypic plasticity is nearly synonymous to acquired changes of a phenotype (as in lifestyle related diseases) that are not accompanied by genetic mutations. Although epigenetic mutation is a broad concept, it will be used synonymously as phenotypic plasticity in the following.

\subsection{Modeling}
In order to simulate Case C above, we formulate the cooperative model as follows. Consider a population consisting of $N$ individuals of a certain species. Each individual has $L$ genes. Let $g(i,j)$ represent the mutation of the gene $j(= 1,\cdots,L)$ of the individual $i (=1,\cdots,N)$, that is, $g(i,j) = 0$ for the wild type, and $g(i,j) = 1$ for a mutant. The contribution $w(j)$ of the gene $j$ to the phenotype may be favorable ($w(j) = 1$) or unfavorable ($w(j) = -1$), and it does not depend on individuals. Individuals embody phenotypic variations due to epigenetic mutations. This epigenetic effect $E(i)$ is assumed to be a Gaussian random  variable with mean 0 and standard deviation $\sigma$. $E(i)> 0$ indicates the epigenetic effect on phenotype is favorable (i.e., adaptive to the environment) and $E(i)<0$, unfavorable. $E(i)$ depends on individuals and is not inherited. The overall change of phenotype, $F(i)$, of individual $i$ is given as
\begin{equation}
F(i) = G(i) + E(i) \label{eq:fitness}
\end{equation}
where 
\begin{equation}
  \label{eq:genetic}
  G(i) = \sum_{j=1}^{L}w(j)g(i,j)
\end{equation}
is the total effect of genetic mutations.
Here, we consider only such phenotypic changes that are involved in natural selection so that we identify $F(i)$ with fitness (the degree of adaptation).

\subsection{Simulation}
Let the population at the time when the environment has changed be the 0-th generation. We now study by computer simulation how the population size evolves as the generation proceeds. We performed two kinds of simulations. In one set of simulations, the fitness depends on both genetic and epigenetic mutations as in Eq. (\ref{eq:fitness}). In the other set, the epigenetic effect is turned off, that is, $E(i) = 0$, so that the fitness depends only on genetic mutations.

In all the simulations, we set the number of genes $L = 10$, $w(j)$ is 1 for 5 genes and -1 for 5 genes, the initial population size $N_0 = 2,000$, the maximum population size $N_m = 100,000$, initial mutation rate $p = 0.05$, random selection rate $q = 0.15$. When the epigenetic effect is absent, the maximum possible value of the fitness $F(i)$ is 5. Thus we set the threshold $T = 5$ for selection. The standard deviation of epigenetic effect was set to $\sigma = 3$.

The procedure of a simulation is the following:
\begin{enumerate}
\item Generate the initial population of $N = N_0$ individuals, each having $L$ (wild type) genes.
\item Mutate each gene in each individual with probability $p$.
\item Select $2N$ pairs out of $N$ individuals and crossover each pair at a randomly selected site. Thus, $2N$ new individuals are produced.
\item For the new individuals, select those with $F(i) > T$ with probability 1, and those with $0 < F(i) < T$ with probability $q$. Those that are not selected are discarded.
\item Set $N$ to the number of remaining individuals. If the remaining population size exceeds $N_m$, randomly select $N_m$ individuals and set $N = N_m$. 
\item Iterate steps (3) to (5).
\end{enumerate}

\subsection{Results}
\begin{figure}[htb]
  \centering
  \includegraphics[width=15cm]{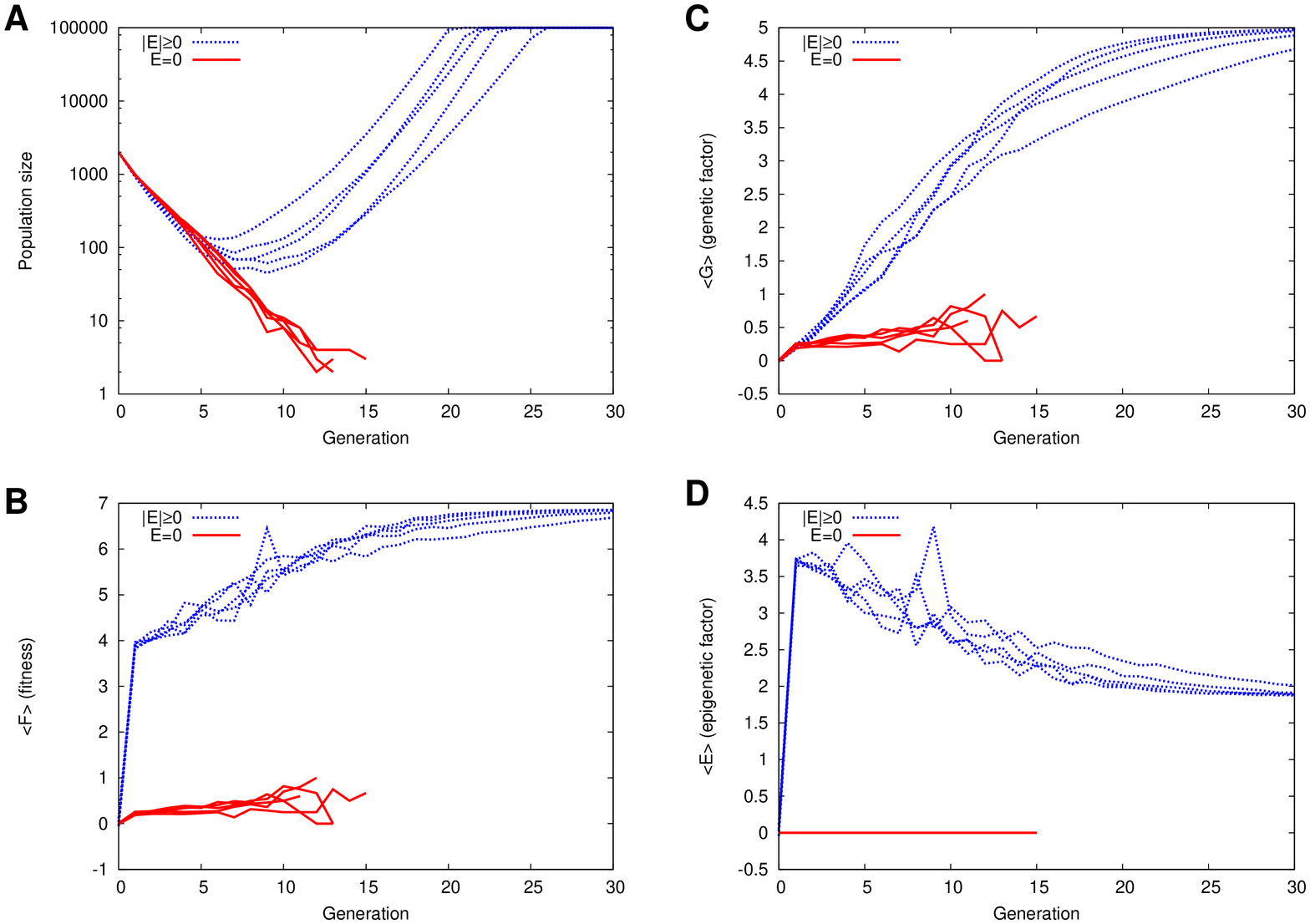}

\caption{{\label{fig:fig1}} Trajectories of the simulations along generations with (blue, $|E|\geq 0$) or without (red, $E = 0$) epigenetic effects. Each type of simulations was run 5 times with different random seeds.
(A) Population size. (B) Average fitness of each population.
(C) Average effect of genetic mutations, $G(i)$, of each population (c.f., Eqs. \ref{eq:fitness} and \ref{eq:genetic}).
(D) Average effect of epigenetic mutation (phenotypic plasticity), $E(i)$, of each population (c.f., Eq. \ref{eq:fitness}).}
\end{figure}

The results of the simulations are shown in Figure 1. When the environment changes, individuals of the standard (wild) type become unstable; those that adapt to the new environment will survive, and those that do not adapt will be eliminated from the population. Figure 1A shows that the initial population size of 2000 rapidly decreases down to below 100 in the first several generations after the environmental change. Up to this point, the two models [with (blue) or without (red) epigenetic effect] behave similarly, but the behaviors become clearly different thereafter. The population of the conventional model with only genetic mutations (red) continues to decrease and becomes extinct after around the 10th generation. On the other hand, the population of the cooperative model (blue) with epigenetic mutations starts to increase at a certain point and continues to grow as generations proceed. Figure 1B indicates that this change of population size is due to the gradual increase of the average fitness of the population, which in turn leads to an increased number of individuals whose fitness surpass the threshold ($T = 5$). At the same time, the average effect of genetic mutations per individual, $\langle G \rangle$ (corresponding to the first term on the right hand side of Eq. \ref{eq:fitness}), is persistently increasing through generations (Figure 1C, blue line). This is an important point, implying that, although the population decreases during the first several generations, adaptive genetic mutations continuously increase on average by genetic recombination. On the contrary, such accumulation of genetic mutations does not occur in the conventional model, and the average effect of genetic mutations remains less than 1. Figure 1D clearly indicates that such a difference between the conventional model and cooperative model is caused by the absence or the presence of the epigenetic factor or phenotypic plasticity (the second term on the right hand side of Eq. \ref{eq:fitness}). 

In these simulations, both the genetic and epigenetic mutations are assumed to occur either favorably [$G(i)> 0$ and/or $E(i)>0$] or unfavorably [$G(i)<0$ and/or $E(i)<0$] with equal probability.
As a result, when an individual has unfavorable genetic or phenotypic mutations, it becomes more likely to be eliminated so that the population as a whole will decrease. It is notable that, even in such a situation, favorable genetic mutations are being accumulated in individuals.

\section{Discussion}
\subsection{Comparison with genetic assimilation}
The genetic assimilation of Waddington assumes that phenotypic plasticity proceeds before genetic mutations in response to a change in environment. This assumption also applies to the cooperative model. In the both cases, no information of phenotype is fed back to genotype so that the prohibition of inheritance of acquired traits is not violated. 

In genetic assimilation, an adaptive phenotypic mutation is maintained solely by the mechanism of phenotypic plasticity until a genetic mutation occurs that realizes the same phenotypic trait. In the cooperative model, genetic assimilation proceeds in a stepwise manner by the cooperation between genetic mutations and 
the phenotypic plasticity rather than being achieved abruptly by a single genetic mutation. In this sense, the cooperative model is a modified version of genetic assimilation. But it should be stressed that this modification makes our model decisively different from the original one.

The greatest advantage of the cooperative model lies at the point that ``genetic assimilation proceeds in a stepwise manner.'' Dawkins called such a mode of evolution the ``cumulative selection.''\cite{Dawkins1986}. For example, if the evolution of a phenotypic trait requires 10 genetic mutations, it is hardly possible to accumulate 10 such mutations in one individual at once, even if mutations can be transferred or exchanged by genetic recombination. However, if each obtained mutation is not lost, more mutations will be accumulated every time a recombination occurs, and as soon as all the mutations are accumulated, the evolution of the phenotypic trait completes. Although Dawkins argues that this mechanism wildly accelerates evolution, he does not provide any explanation as to how the cumulative selection is possible within the framework of the Darwinian evolution that is based solely on genetic mutations and natural selection. After all, what is important for the cumulative selection is some kind of a ``ratchet'' mechanism by which intermediate steps of accumulating genetic mutations are maintained. In our model, the ratchet mechanism operates in such a way that individuals are selected when the sum of genetic and epigenetic effects surpasses a threshold. Figure 1 indeed indicates that successful survival strongly depends on whether the cumulative selection is present (blue) or not (red).

\subsection{Reexamining assumptions}
Let us now examine several (possibly implicit) assumptions in the cooperative model.

\begin{itemize}
\item Natural selection operates on phenotypes of individuals (not on genotypes).
\end{itemize}
In the conventional theory of evolutionary synthesis, ontogenesis is either ignored or treated as a black box so that it effectively equates phenotype with genotype (phenotype = genotype). In other words, if we ignore developmental process, it does not matter if natural selection acts on genotypes or on phenotypes.
For example, in textbooks of population genetics and molecular evolution, we often find such a statement as ``Evolution of the population of a species is a process of the temporary change of allele frequencies due to natural selection'' \cite{Saitou2006}. Here, no mention is found as to not only ontogenesis but even phenotype. However, we cannot ignore ontogenesis so that such a view of evolution is simply incorrect. 
One conspicuous example of the plasticity of development is the various (nearly 200) types of human cells. Although all the somatic cells have the same genomic information, a diverse set of cells (epidermis, liver, muscle, neuron, etc.) are expressed through cell differentiation. This variety may be regarded as the range of possible acquired changes of the cell. In practice, the process of cell differentiation is strictly controlled by epigenetic programs. Nevertheless, if there is a small change in such a program, the phenotype may easily change. Gilbert and Epel\cite{GilbertANDEpel2008} enumerates many examples of developmental (or phenotypic) plasticity in response to environmental conditions. Natural selection should be considered to act on phenotypes of individuals including such changes of developmental process.

\begin{itemize}
\item Phenotypic changes (new traits) are in general obtained by genetic factors and phenotypic plasticity.
\item The relationship between traits (phenotypes) and genotypes is not one-to-one, but one-to-many.
\end{itemize}
We omit detailed discussion regarding these two points since they are already mentioned above in relation to lifestyle related diseases.
A more fundamental assumption is the following.

\begin{itemize}
\item Sexual reproduction.
\end{itemize}
Genetic recombination that is followed by mating and fertilization is an indispensable assumption for the cooperative model. As shown by the simulations, accumulation of genetic mutations is made possible by genetic recombination. Therefore, the cooperative model is applicable to most eukaryotic organisms, but not to prokaryotes. In unicellular prokaryotic organisms such as bacteria, cells are small and the ``distance'' between genotype and phenotype is also small so that one may assume genotypes being nearly equal to phenotypes. Then, the conventional theory of evolutionary synthesis is more appropriate for treating the evolution of bacteria. On the other hand, eukaryotic organisms have larger cells and each cell is compartmentalized by membrane systems comprising internal structures such as nucleus, mitochondria, and other organella. Accordingly, the distance between genotype and phenotype becomes larger, and the distance is even larger in multicellular organisms such as plants and animals. The large distance from the genotype to the macroscopic phenotype implies that there are many layers of control which in turn provide the freedom or plasticity of ontogenesis. By the mechanism of the cooperation between developmental (phenotypic) plasticity and genetic mutations, Dawkins' cumulative selection operates, and the evolution of multicellular organisms may have been accelerated.
This argument suggests that sexual reproduction is superior to asexual one in accelerating biological evolution.

\begin{itemize}
\item Evolution begins with an environmental change.
\end{itemize}
This model assumes that a change of environment is necessary for evolution to occur. In other words, organisms are assumed to be passive to the environment and to remain status quo as long as the environment stays constant. 
For example, if climate changes over yearly periods (such as in ice age), organisms will (reluctantly) adapt to the new climate, otherwise they will be extinct. The wild type species that have been stable under a certain environment will not be stable in the changed environment so that they will either evolve to adapt or become extinct \cite{Yoshimura2009}. However natural this idea is, the effect of environmental changes is, surprisingly, hardly treated in the conventional theory of evolutionary synthesis. For example, in Mayr's \emph{Toward a New Philosophy of Biology}\cite{Mayr1988}, a representative textbook of evolutionary synthesis, there is no such statement as ``environmental changes drive evolution'' in the text and there are no such terms as ``environment'' or ``environmental change'' in the index. This mysterious convention of evolutionary biologists probably dates back to Darwin. Darwin's explanation of evolution starts from mutations of individuals. \emph{The Origin of Species} discusses domesticated animals and selective breeding from the beginning.
Indeed, artificial selection of domesticated animals serves as the model system for natural selection, and this approach inevitably starts from mutants rather than environmental changes. By today's standard, it is fair to say that Darwin, who was not aware of continental drift and global mass extinction, was constrained by his time.

\subsection{Restoration of the law of use and disuse}
If the above assumptions are all valid, or at least plausible, the simulations shown in Figure 1 will be more reliable. If the cooperative model is correct, it is necessary to reexamine some problems that have been considered to be taboos in Darwinian theory of evolution.

Our model of evolution can be summarized as ``phenotypic traits (phenotypes) induced by phenotypic plasticity may be genetically stabilized.'' Thus summarized, this model appears very similar to the law of use and disuse proposed by Lamarck. While Lamarck assumed that the acquired phenotypic change was inherited, the present model assumes such is not possible. Thus the present model leads to the following hypothesis: acquired traits are not inherited, but some of them will be genetically fixed, or genetically assimilated, after several generations.

To give an example, a classical problem pointed out by Waddington is why ostriches form calluses at the right place (rumps) in the right manner \cite{GilbertANDEpel2008}. This is a very hard problem if we try to solve it based on the conventional evolutionary synthesis. However, it can be solved rather easily if we employ the law of use and disuse. When an ostrich sits down, its rumps touch the ground, and they will be hardened and form calluses. This is a normal response (phenotypic plasticity) to physical stresses imposed on skin. 
Sitting on the ground is important to ostriches, and in the beginning, it might have hurt because of the rubbed calluses, but they cannot avoid but sitting down. After some generations, favorable, adaptive mutations are accumulated, and genetically fixed (the cooperative model). In fact, ostriches' calluses are a genetic trait and they start to form during embryogenesis \cite{Gould1980}.

Although the law of use and disuse practically holds, it does not mean that all the acquired traits (phenotypic mutations) can be genetically fixed. For example, massive muscles of a blacksmith are not inherited to his descendants. This is because massive muscles are not absolutely necessary for humans to survive and hence not subject to natural selection. The same argument applies to body height and weight. These are not subject to natural selection although there are individual variations. 

To animals, securing foods is an elementary requisite for survival, especially the main diet of each animal species is decisively important. Even only within primates, the wide variety of palates and hands is astonishing, 
their diverse shapes are considered to be due to the difference of their main diet \cite{Shima2003}.
For example, chimpanzees that mainly eat fruits have sharp cuspids, supposedly to tear the hard outer peels of fruits.
In comparison, human teeth are rather flat and cuspids are small. The shapes of hands and fingers of primates is bewildering, one of the most conspicuous of them is perhaps those of aye-ayes (\emph{Daubentonia madagascariensis}).
Aye-ayes have thin and long middle fingers which serve for two purposes: ``tapping to find insects within cavities in wood and probing to find and remove larvae from those cavities'' \cite{Gron2007}.
Such examples of adaptive evolution are numerous.

In the late last century, some observations of biological evolution have been made in the fields. Among them, Darwin's finches are famous for being intensively investigated by Rosemary and Peter Grant and their colleagues for about 30 years \cite{Weiner1994}. According to their study, the phenotype of the shape (round or sharp) and size of the beak of Darwin's finches changes due to environmental changes such as heavy rains or droughts. One year, a drought hit the island, resulting in the decreased number of small soft seeds which were the main diet of the finches. Instead, large and hard fruits of cactus spread. Then, the relative reproduction rate of the finches with large round beaks increased, and later offspring tended to have such beaks. 10 years later, heavy rains and high temperature caused by El Ni{\~n}o decreased the number of cactus, and plants with small soft seeds spread. This lead to the increase of relative reproduction rate of finches with small sharp beaks and the ratio of such finches increased. These yearly changes of finches' phenotypes largely exceed the range of random variations (genetic drift), and may be regarded as evolution by natural selection \cite{Kawata2006}. It is noted that the evolution of Darwin's finches started with environmental changes and that such evolutionary changes are unexpectedly rapid (at least twice in 30 years). These observations are consistent with the cooperative model.

\subsection{Bridging between ontogenesis and evolution}
The phenotype-driven mechanism of evolution can be regarded as an evolutionary mechanism based on ontogenesis. Gilbert and Epel \cite{GilbertANDEpel2008} summarized the advantages of ontogenetic evolutionary models:
(1) ``Mutationally induced novelties would occur in only a family of individuals, whereas environmentally induced novelties would occur throughout a population''; (2) ``The inducing environment would most often also be a selecting environment.'' 

These points indicate that the coupling between an environmental change and phenotypic plasticity induces phenotypic mutants in many individuals in the population simultaneously and promotes an evolutionary change in the population as a whole, which leads to an accelerated evolution.

Furthermore, Gilbert and Epel \cite{GilbertANDEpel2008} seem to consider the possibility that epigenetic changes may be directly inherited to next generations through the germ line. There are some experimental evidences suggesting that it is possible, but that does not necessarily imply that the Weismann barrier is broken. Even if epigenetic changes may be (partially) inherited, that does not imply a change in genotype (i.e., genome sequence). Evolution must be accompanied by some change of genotype so that (inherited) epigenetic changes must be somehow replaced with genetic mutants. But Gilbert and Epel do not show how such replacement of epigenetic changes with genetic mutations is possible. If we regard epigenetic changes as a kind of phenotypic plasticity, 
it is possible to incorporate epigenetic effects into the cooperative model, and thus epigenetic changes can be linked to evolution.

\emph{Developmental plasticity and evolution} by West-Eberhard, as the title itself suggests, is a huge treatise on ontogenesis and evolution. Although it is not easy to summarize, some of the main points are the following.
\begin{enumerate}
\item Environmental changes induce phenotypic mutations.
\item The primary determinant of an organism's design (phenotype from a functional view point) is the effect of environment on ontogenesis, not genetic mutations.
\item Phenotypic plasticity proceeds and is gradually replaced with genetic mutations.
\item Quantitative genetic mutations are involved in phenotypic mutations expressed in the process of ontogenesis.
\item Evolutionary developmental biology demonstrated that gene sets in genomes are well conserved beyond taxonomic groups.
\end{enumerate}
These statements are not compatible with the conventional theory of evolutionary synthesis, but are consistent with our cooperative model.

In their \emph{Plausibility of Life: Resolving Darwin's Dilemma} \cite{KirschnerANDGerhart2005}, Kirschner and Gerhart starts with the question as to how the accumulation of random mutations leads to an adaptive new traits. They cast this problem as the relationship between genotypes and phenotypes, and examine recent findings of animal developmental biology based on fundamentals of molecular and cellular biology. Their explanation on the process of developmental pattern formation based on ``sections'' of embryo is noteworthy. They argue that new traits emerge when elementary cellular processes and developmental processes are ``deconstrained,'' and propose that such deconstraining is promoted by genetic mutation (theory of facilitated phenotypic variation). However, there is a serious flaw in this theory, for it states that ``random genetic mutations induces an adaptive phenotype through ontogenesis,'' which is too opportunistic (random genetic mutations affecting developmental process should also induce unfavorable phenotypes).
To correct their theory, random genetic mutations and deconstraining of developmental process should not be connected by a causal relationship. Then, we should assume that the cooperation between developmental plasticity and genetic mutations induces a phenotypic mutation on which natural selection operates, and hence evolution proceeds (the cooperative model). In this manner, ontogenesis and evolution can be seamlessly and logically connected.

\section{Conclusion}
As we have discussed, there are many models based on the phenotype-driven mechanism of evolution. As in evolutionary developmental biology, it is often proposed that evolution and development should be somehow integrated\cite{Carroll2005}. 
To date, however, no theory seems to have succeeded in establishing the ``true evolutionary synthesis'' that overcomes the difficulties of the conventional evolutionary synthesis. If we pursue the reason for this difficulty, we find the Weismann barrier standing invulnerable. It is not difficult to imagine that phenotypic mutations induced by developmental plasticity is important and it plays some pivotal role in evolution. At this point, we are faced with the fact that it is prohibited to feed back the acquired phenotypic mutations to genotype. But if we directly equate the phenotype to the genotype to circumvent the difficulty, the theory will inevitably become logically inconsistent, as we have seen in the present work. The ``cooperative model'' proposed in this paper is a small modification of ``genetic assimilation'' or ``theory of facilitated phenotypic variation.'' 
Nevertheless, we believe that there is a significant implication in this small modification. Theory of evolution has been based on two components, genetics and evolution, but it must be extended to include ontogenesis. Such three-component theory of evolution, when established, should be properly called the ``true theory of evolutionary synthesis.'' We hope that our model serves as a step toward such a theory.

\section*{Acknowledgements}
We thank Drs. Haruki Nakamura, Takeshi Kawabata, and Toshiyuki Takano-Shimizu for assessing the manuscript.

%\bibliographystyle{biophysics}
%\bibliography{kn}

%\section*{Figure Legend}
% \begin{figure}[htb]
%   \centering
%   \includegraphics[width=15cm]{fig1}

% \caption{{\label{fig:fig1}} Trajectories of the simulations along generations with (blue, $|E|\geq 0$) or without (red, $E = 0$) epigenetic effects. Each type of simulations was run 5 times with different random seeds.
% (A) Population size. (B) Average fitness of each population.
% (C) Average effect of genetic mutations, $G(i)$, of each population (c.f., Eqs. \ref{eq:fitness} and \ref{eq:genetic}).
% (D) Average effect of epigenetic mutation (phenotypic plasticity), $E(i)$, of each population (c.f., Eq. \ref{eq:fitness}).}
% \end{figure}

\end{document}